\documentclass[10pt,conference]{IEEEtran}
\usepackage{amssymb}
\usepackage{amsbsy}
\usepackage{amsmath}
\usepackage{amsthm} % This is only for theorem and proofs.
\usepackage{setspace}
\usepackage{epsfig}
\usepackage{cite}
\usepackage{graphics}
\usepackage{epstopdf}
\usepackage{mathrsfs}
\usepackage[above]{placeins}
\usepackage{algorithm}
\usepackage{algpseudocode}
\usepackage{multirow}
\usepackage{algorithmicx}
\usepackage{tikz}

\begin{document}
\title{Markov Chain Monte Carlo Algorithms \\
for Lattice Gaussian Sampling}
%Department of Electrical and Electronic Engineering
\author{\IEEEauthorblockN{Zheng Wang and Cong Ling}
\IEEEauthorblockA{Department of EEE\\ Imperial College London\\
London, SW7 2AZ, United Kingdom\\
Email: z.wang10, c.ling@imperial.ac.uk}
\and
\IEEEauthorblockN{Guillaume Hanrot}
\IEEEauthorblockA{\'{E}cole Normale Sup\'{e}rieure de Lyon\\
LIP (CNRS, ENS Lyon, UCBL, INRIA) \\
46 All\'{e}e d'Italie, 69364 Lyon Cedex 07, France\\
Email:Guillaume.Hanrot@ens-lyon.fr}}
%Department of Computer Science

\maketitle

\begin{abstract}
To be considered for an IEEE Jack Keil Wolf ISIT Student Paper Award.

Sampling from a lattice Gaussian distribution is emerging as an important problem in various areas such as coding and cryptography. The default sampling algorithm --- Klein's algorithm yields a distribution close to the lattice Gaussian only if the standard deviation is sufficiently large. In this paper, we propose the Markov chain Monte Carlo (MCMC) method for lattice Gaussian sampling when this condition is not satisfied. In particular, we present a sampling algorithm based on Gibbs sampling, which converges to the target lattice Gaussian distribution for any value of the standard deviation. To improve the convergence rate, a more efficient algorithm referred to as \emph{Gibbs-Klein sampling} is proposed, which samples block by block using Klein's algorithm. We show that Gibbs-Klein sampling yields a distribution close to the target lattice Gaussian, under a less stringent condition than that of the original Klein algorithm.
\end{abstract}

%\begin{IEEEkeywords}
%Lattice Gaussian, Gibbs sampling, lattice sampling, Markov Chain Monte Carlo.
%\end{IEEEkeywords}

% \fi
\IEEEpeerreviewmaketitle

\section{Introduction}
The lattice Gaussian distribution is emerging as a common theme in various areas. In mathematics, Banaszczyk \cite{Banaszczyk} firstly used it to prove the transference theorems of lattices. In coding, it mimics Shannon's Gaussian random coding technique, yet permits lattice decoding. Forney applied the lattice Gaussian distribution to obtain the full shaping gain in lattice coding \cite{Forney_89} (see also \cite{Kschischang_Pasupathy}). Recently, it has been used to achieve the capacity of the Gaussian channel \cite{LB_13} and to approach the secrecy capacity of the Gaussian wiretap channel \cite{LLBS_12}, respectively. Sampling from the lattice Gaussian has also been used in lattice decoding for the multi-input multi-output system \cite{CongRandom,DerandomizedJ}. In cryptography, lattice Gaussians have become a central tool in the construction of many primitives. Micciancio and Regev used it to propose lattice-based cryptosystems based on the worst-case hardness assumptions \cite{MicciancioGaussian}, and recently, it has underpinned the fully-homomorphic encryption for cloud computing \cite{GentrySW13}. The key fact is again that a vector distributed as a lattice Gaussian centered at $\mathbf{c}$ with a small standard deviation is typically very close to $\mathbf{c}$. To illustrate why this
might be useful in cryptography, note that if one knows a short basis
of the lattice, one can efficiently produce such a vector \cite{Trapdoor},
while disclosing no information on the short basis---since the
lattice Gaussian distribution does not depend on the particular basis.

Thus, in both coding and cryptography, efficient sampling algorithms for the lattice Gaussian as well as a good understanding on how the complexity depends on the standard deviation is an important issue. However, in contrast to sampling from the continuous Gaussian distribution, it is not at all straightforward to sample from a discrete Gaussian distribution over a lattice. At present, the default sampling algorithm for lattices is due to Klein, originally proposed for bounded-distance decoding \cite{Klein} (see also \cite{Peikert10,BrakerskiLPRS13} for variations and \cite{LB_13} for an algorithm for lattices of Construction A). It was shown in \cite{Trapdoor} that Klein's algorithm samples within a negligible statistical distance from the lattice Gaussian distribution only if the standard deviation $\sigma \geq \omega(\sqrt{\text{log}\ n})\cdot\text{max}_{1\leq i \leq n}\|\mathbf{\widehat{b}}_i\|$, where $n$ is the lattice dimension and $\mathbf{\widehat{b}}_i$'s are the Gram-Schmidt vectors of the lattice basis. Unfortunately, such a requirement of $\sigma$ can be excessive, rendering Klein's algorithm inapplicable to many cases of interest.

Markov chain Monte Carlo (MCMC) methods attempt to sample from the target distribution of interest by building a Markov chain, which randomly generate the next sample conditioned on the previous samples. As a major algorithm of MCMC, Gibbs sampling \cite{LiuBook} constructs a Markov chain which gradually converges to the target distribution by only considering univariate sampling at each step. In this paper, we introduce the Gibbs algorithm into lattice Gaussian sampling and propose a more efficient block-based algorithm named as \emph{Gibbs-Klein sampling}. In contrast to conventional blocked sampling which is computationally more demanding, the proposed algorithm takes advantages of Klein's algorithm as a building block. The proposed algorithms are applicable in the scenario $\sigma < \omega(\sqrt{\text{log}\ n})\cdot\text{max}_{1\leq i \leq n}\|\mathbf{\widehat{b}}_i\|$.

To the best of our knowledge, this is the first time that MCMC methods are used in lattice Gaussian distributions. Different from previous works on Gibbs sampling for signal detection of finite constellations \cite{MCMCBehrouz,HassibiMCMC,ChenMCMC}, here we are concerned with countably infinite state spaces and with simulating Gaussian distributions over a lattice. It is worth pointing out that although the underlying Markov chain converges to the stationary distribution for all values of $\sigma$, the convergence is
expected to become very slow when $\sigma$ becomes small, since for very small
$\sigma$ we would solve the closest vector problem (CVP) and shortest vector problem (SVP) with high probability.
%rather than systematic-scan version with finite constellation $\mathcal{X}^n$

%Unfortunately, despite 60 years of research, any attempt to assess the convergence rate or mixing time of an MCMC sampler is still formidable even for some simple-like problems, leaving a big open question in the area of MCMC. Therefore, in this paper such unsolved issues are not involved and we restrict ourselves to focus comparisons of the convergence rate rather than exact convergence rate assessment.

%In spite of the depressed fact of the convergence rate estimation, the mixing time that the proposed Gibbs-Klein sampling needs to step into the stationary distribution of Markov chain

The rest of this paper is organized as follows. Section II introduces
lattice Gaussian distributions and briefly reviews Klein's algorithm.
In Section III, the conventional Gibbs and the new Gibbs-Klein sampling algorithms are proposed for lattice Gaussians, followed by a theoretical analysis in Section IV. Section V presents the simulation results.

%Klein's algorithm was firstly proposed in \cite{Klein} to solve the closest vector problem (CVP) in lattice decoding. Compared with Babai's nearest plane algorithm (also known as successive interference cancelation) who performs the detection resorting to direct rounding, Klein's algorithm obtains the detected signal by random sampling over a conditioned 1-dimensional Gaussian distribution.

\section{Lattice Gaussian Distributions}

%\subsection{Lattice Gaussian Distribution}
Let $\mathbf{B}=[\mathbf{b}_1,\ldots,\mathbf{b}_n]\subset \mathbb{R}^n$ consist of $n$ linearly independent vectors. The $n$-dimensional lattice $\Lambda$ based on $\mathbf{B}$ is defined by
\begin{equation}
\Lambda=\mathcal{L}(\mathbf{B})=\{\mathbf{Bx}: \mathbf{x}\in \mathbb{Z}^n\},
\end{equation}
where $\mathbf{B}$ is known as the lattice basis. We define the Gaussian function centered at $\mathbf{c}\in \mathbb{R}^n$ for standard deviation $\sigma>0$ as
\begin{equation}
\rho_{\sigma, \mathbf{c}}(\mathbf{z})=e^{-\frac{\|\mathbf{z}-\mathbf{c}\|^2}{2\sigma^2}},
\end{equation}
%\begin{equation}
%\rho_{\sigma, \mathbf{c}}(\mathbf{z})=\frac{1}{(\sqrt{2\pi}\sigma)^n}e^{-\frac{\|\mathbf{z}-\mathbf{c}\|^2}{22\sigma^2}},
%\end{equation}
for all $\mathbf{z}\in\mathbb{R}^n$. Then, the \emph{discrete Gaussian distribution} over $\Lambda$ is defined as
\begin{equation}
D_{\Lambda,\sigma,\mathbf{c}}(\mathbf{x})=\frac{\rho_{\sigma, \mathbf{c}}(\mathbf{Bx})}{\rho_{\sigma, \mathbf{c}}(\Lambda)}=\frac{e^{-\frac{1}{2\sigma^2}\parallel \mathbf{Bx}-\mathbf{c} \parallel^2}}{\sum_{\mathbf{x} \in \mathbb{Z}^n}e^{-\frac{1}{2\sigma^2}\parallel \mathbf{Bx}-\mathbf{c} \parallel^2}}
\end{equation}
for all $\mathbf{Bx}\in \Lambda$, where $\rho_{\sigma, \mathbf{c}}(\Lambda)\triangleq \sum_{\mathbf{\mathbf{Bx}}\in\Lambda}\rho_{\sigma, \mathbf{c}}(\mathbf{Bx})$.

An intuition of $D_{\Lambda,\sigma,\mathbf{c}}(\mathbf{x})$ suggests that the closer lattice point $\mathbf{Bx}$ is to $\mathbf{c}$, the higher probability it will be sampled. Thus, lattice Gaussian sampling can be applied to solve the CVP, and Klein's algorithm was originally proposed for decoding \cite{Klein}. As a randomized version of Babai's nearest-plane algorithm (i.e., successive interference cancellation), Klein's algorithm obtains a vector by sequentially sampling from a 1-dimensional conditional Gaussian distribution. As shown in Algorithm 1, its operation has polynomial complexity $O(n^2)$ excluding QR decomposition.
\renewcommand{\algorithmicrequire}{\textbf{Input:}}  % Use Input in the format of Algorithm
\renewcommand{\algorithmicensure}{\textbf{Output:}} % Use Output in the format of Algorithm

%\begin{algorithm}[h]
%\caption{Klein's Algorithm}
%\begin{algorithmic}[1]
%\Require
%$\mathbf{B}, \sigma, \mathbf{c}$;
%\Ensure
%$\mathbf{Bx}$;
%\State let $\mathbf{v}_n=\mathbf{0}$ and $\mathbf{c}_n=\mathbf{c}$
%\For {$i=n$,\ \ldots,\ 1}
%\State let $c^{'}_i=\langle\mathbf{c}_i, \mathbf{\widehat{b}}_i\rangle/\langle\mathbf{\widehat{b}}_i,\mathbf{\widehat{b}}_i\rangle\in\mathbb{R}$ and $\sigma^{'}_i=\sigma/\|\mathbf{\widehat{b}}_i\|$
%\State sample $\widehat{x}_i$ from $D_{\mathbb{Z},\sigma^{'}_i,c^{'}_i}$
%\State let $\mathbf{c}_{i-1}=\mathbf{c}_{i}-\mathbf{b}_i\widehat{x}_i$ and $\mathbf{v}_{i-1}=\mathbf{v}_{i}+\mathbf{b}_i\widehat{x}_i$
%\EndFor
%\State return $\mathbf{Bx}=\mathbf{v}_0$
%\end{algorithmic}
%\end{algorithm}

\begin{algorithm}[t]
\caption{Klein's Algorithm}
\begin{algorithmic}[1]
\Require
$\mathbf{B}, \sigma, \mathbf{c}$
\Ensure
$\mathbf{Bx}\in\Lambda$
\State let $\mathbf{B}=\mathbf{QR}$ and $\mathbf{c'}=\mathbf{Q}^{T}\mathbf{c}$
\For {$i=n$,\ \ldots,\ 1}
\State let $\alpha_i=\frac{\sigma}{|r_{i,i}|}$ and $\widetilde{x}_i=\frac{c'_i-\sum^n_{j=i+1}r_{i,j}x_j}{r_{i,i}}$
\State sample $x_i$ from $D_{\mathbb{Z},\alpha_i,\widetilde{x}_i}$
\EndFor
\State return $\mathbf{Bx}$
\end{algorithmic}
\end{algorithm}

The parameter $\sigma$ is key to the distribution produced by Klein's algorithm. Klein suggested  $\sigma=\text{min}_i\|\mathbf{\widehat{b}}_i\|/\sqrt{\text{log}\ n}$ and this was followed/adapted in \cite{CongRandom,DerandomizedJ}. In this case, Klein's algorithm only yields a distribution that is lower-bounded by the Gaussian distribution. On the other hand, it was demonstrated in \cite{Trapdoor} that Klein's algorithm actually samples from $D_{\Lambda,\sigma,\mathbf{c}}$ within a negligible statistical distance if
\begin{equation}
\sigma\geq\omega(\sqrt{\text{log}\ n})\cdot\text{max}_{1\leq i \leq n}\|\mathbf{\widehat{b}}_i\|.
\end{equation}
However, Gaussian sampling algorithms are lacking for the range $\sigma<\omega(\sqrt{\text{log}\ n})\cdot\text{max}_i\|\mathbf{\widehat{b}}_i\|$.

%It is necessary to point out that the original $\sigma$ Klein chose maintains $\sigma=\text{min}_i\|\mathbf{\widehat{b}}_i\|/\sqrt{\text{log}\hspace{0.1em}n}$ and then the running times of Klein's algorithm of returning the closest lattice point to the given point could be further derived as $O(n^{\|\lambda-\mathbf{c}\|^2/\text{min}_i\|\mathbf{\widehat{b}}_i\|})$. Apparently, if the distance from $\mathbf{c}$ to the lattice is not much more than the length of the smallest Gram-Schmidt vector, Klein's algorithm will tend to terminate within polynomial complexity $O(n^{\|\lambda-\mathbf{c}\|^2/\text{min}_i\|\mathbf{\widehat{b}}_i\|+2})$.

%\subsection{Klein's Sampling Algorithm}
%Klein's algorithm was firstly proposed in \cite{Klein} to solve the closest vector problem (CVP) in lattice decoding. Compared with Babai's nearest plane algorithm (also known as successive interference cancelation) who performs the detection resorting to direct rounding, Klein's algorithm obtains the detected signal by random sampling over a conditioned 1-dimensional Gaussian distribution.

%$D_{\mathbb{Z},\sigma^{'}_i,c^{'}_i}$

%In other words, the way of realizing that discrete Gaussian distribution

\section{MCMC for Lattice Gaussian}
In this section, we introduce the concept of MCMC into lattice Gaussian sampling for the range of $\sigma$ where Klein's algorithm cannot reach. We further propose a more efficient sampling algorithm named as \emph{Gibbs-Klein sampling} to improve the convergence rate.

\subsection{Gibbs Sampling for Lattice Gaussian}
Lattice Gaussian distribution $D_{\Lambda,\sigma,\mathbf{c}}$ with $\sigma<\omega(\sqrt{\text{log}\ n})\cdot\text{max}_i\|\mathbf{\widehat{b}}_i\|$ can be seen as a complex target distribution lacking direct sampling methods. MCMC makes use of the conditional distribution as a tractable alternative to work with. Here we apply the Gibbs algorithm to sample from the original joint distribution $D_{\Lambda,\sigma,\mathbf{c}}$.

Gibbs sampling employs 1-dimensional conditional distributions to construct the Markov chain \cite{LiuBook}, where all other variables in the distribution are unchanged in each step. In this way, we sample $n$ random variables from the corresponding $n$ univariate conditionals in a certain order instead of directly generating an $n$-dimensional vector. Samples drawn from the target joint distribution will be generated when the Markov chain reaches the stationary distribution.

Specifically, in Gibbs sampling, each coordinate of $\mathbf{x}$
is sampled from the following 1-dimensional conditional distribution
\begin{eqnarray}
P(x_i^{t+1}|\mathbf{x}_{[-i]}^t)&=&\frac{e^{-\frac{1}{2\sigma^2}\parallel \mathbf{Bx}^{t+1}-\mathbf{c} \parallel^2}}{\sum_{x_i^{t+1} \in \mathbb{Z}}e^{-\frac{1}{2\sigma^2}\parallel \mathbf{Bx}^{t+1}-\mathbf{c} \parallel^2}},
\end{eqnarray}
%\begin{eqnarray}
%P(x_i^{l+1}|\mathbf{x}_{[-i]}^l)\hspace{-0.8em}&=&\hspace{-0.8em}\frac{e^{-\frac{1}{2\sigma^2}\parallel \mathbf{Bx-\mathbf{c}} \parallel^2}}{\sum_{x_i^{l+1} \in \mathbb{Z}}e^{-\frac{1}{2\sigma^2}\parallel \mathbf{Bx-\mathbf{c}} \parallel^2}}\notag \\
%\hspace{-0.8em}&=&\hspace{-0.8em}\frac{e^{-\frac{1}{2\sigma^2}\parallel \mathbf{c}^{'}-\mathbf{Rz} \parallel^2}}{\sum_{z_1^{l+1} \in \mathbb{Z}}e^{-\frac{1}{2\sigma^2}\parallel \mathbf{c}^{'}-\mathbf{Rz} \parallel^2}} \notag\\
%\hspace{-0.8em}&=&\hspace{-0.8em}\frac{e^{-\frac{r_{1,1}^2}{2\sigma^2}\left(\frac{c'_1-\sum^n_{j=2}r_{1,j}z_j^l}{r_{1,1}}-z_1^{l+1}\right)^2}}{\sum_{z_1^{l+1}\in \mathbb{Z}}e^{-\frac{r_{1,1}^2}{2\sigma^2}\left(\frac{c'_1-\sum^n_{j=2}r_{1,j}z_j^l}{r_{1,1}}-z_1^{l+1}\right)^2}}\notag\\
%\hspace{-0.8em}&=&\hspace{-0.8em}\frac{e^{-\frac{1}{\alpha^2}\left(\widetilde{z}_1^l-z_1^{l+1}\right)^2}}{\sum_{z_1^{l+1}\in \mathbb{Z}}e^{-\frac{1}{\alpha^2}\left(\widetilde{z}_1^l-z_1^{l+1}\right)^2}},
%\label{1 conditional distribution}
%\end{eqnarray}
where $1\leq i\leq n$ denotes the coordinate index of $\mathbf{x}$, $\mathbf{x}_{[-i]}^t\triangleq[x_1^t,\ldots,x_{i-1}^t,x_{i+1}^t,\ldots,x_{n}^t]^T$, and $t$ is the time index of the Markov chain. It is noteworthy that there are many scan schemes in Gibbs sampling and we apply the random-scan in this paper, which means the index $i$ is randomly chosen at each step. The extension to other scan strategies is possible.
%the other $n-1$ coordinates except $x_i$

By repeating such a procedure, an underlying Markov chain ${\mathbf{x}^{t+1}=[x_1^{t},\ldots,x_{i-1}^{t},x^{t+1}_i,x_{i+1}^{t},\ldots,x_{n}^{t}]^T}$ is induced, whose transition probability between two adjacent states is defined by the univariate Gibbs sampler,
\begin{equation}\label{eq:transition}
 P(\mathbf{x}^t; \mathbf{x}^{t+1})=P(x_i^{t+1}|\mathbf{x}_{[-i]}^t).
\end{equation}
Clearly, every two adjacent states of $\mathbf{x}$ differ from each other by only one coordinate and it is easy to see that $D_{\Lambda,\sigma,\mathbf{c}}$ stays invariant under such transitions. Algorithm 2 gives the operation of Gibbs sampling for lattice Gaussian distributions. The initial random variable $\mathbf{x}^0$ can be chosen from $\mathbb{Z}^n$ arbitrarily or from the output of a suboptimal algorithm, while the time bound $T$ is large enough to reach the stationary distribution $D_{\Lambda,\sigma,\mathbf{c}}$.
\begin{algorithm}[t]
\caption{Gibbs sampling for lattice Gaussian}
\begin{algorithmic}[1]
\Require
$\mathbf{B}, \sigma, \mathbf{c}, \mathbf{x}^0$
\Ensure
$\mathbf{x} \thicksim D_{\Lambda,\sigma,\mathbf{c}}$ as $T\to \infty$
\For {$t=$1,\ \ldots,\ $T$}
\State randomly choose coordinate index $i$ from $\{1,2,\ldots,n\}$
\State sample $x_i$ from $P(x_i^{t}|\mathbf{x}_{[-i]}^{t-1})$
\State update $\mathbf{x}^{t}=[x_1^{t-1},\ldots,x_{i-1}^{t-1},x_i,x_{i+1}^{t-1},\ldots,x_{n}^{t-1}]^T$
\If {Markov chain has reached stationarity}
\State output $\mathbf{x}^{t}$
\EndIf
\EndFor
\end{algorithmic}
\end{algorithm}

%with transition probability
%\begin{equation}
% P(\mathbf{x}^{t+1}|\mathbf{x}^t)=\frac{e^{-\frac{1}{2\sigma^2}\parallel \mathbf{Bx}^{t+1}-\mathbf{c} \parallel^2}}{\sum_{x_i^{t+1} \in \mathbb{Z}}e^{-\frac{1}{2\sigma^2}\parallel \mathbf{Bx}^{t+1}-\mathbf{c} \parallel^2}}.
%\end{equation}
With the transition probabilities (\ref{eq:transition}), we may form the infinite transition matrix $\mathbf{P}$, whose $(i,j)$-th entry $P(s_i; s_j)$ represents the probability of transferring to state $s_j$ from the previous state $s_i$. Denote by $\mathbf{P}^t$ the transition matrix after $t$ steps. We group in the following
theorem standard results about Gibbs sampling \cite{mixingtimemarkovchain}.

%It is routine to check that the Markov chain induced by the Gibbs sampler is
%Based on an irreducible, aperiodic and reversible Markov chain, Theorem 1 shown below is well known and easy to derive \cite{mixingtimemarkovchain}  so that we will skip the proof directly.
%as an unique measure of the Markov chain.
\newtheorem{my1}{Theorem}
\newtheorem{my3}{Proposition}
\begin{my3}
Given the invariant distribution $D_{\Lambda,\sigma,\mathbf{c}}$, the Markov chain induced by the Gibbs sampler is irreducible, aperiodic and reversible (hence positive recurrent), and converges to the stationary distribution in the total variation (TV) distance as $t\to \infty$:
\end{my3}
\vspace{-1em}
\begin{equation}
\underset{t\rightarrow\infty}{\text{lim}}\|P^t(\mathbf{x}; \cdot)-D_{\Lambda,\sigma,\mathbf{c}}\|_{TV}=0,
\end{equation}
\emph{for all states $\mathbf{x}\in\mathbb{Z}^n$, where $P^t(\mathbf{x}; \cdot)$ denotes the row of $\mathbf{P}^t$ corresponding to initial state $\mathbf{x}$.}

According to Proposition 1, if time permits to reach the stationary distribution, the proposed Gibbs sampler will draw samples from $D_{\Lambda,\sigma,\mathbf{c}}$ no matter what value $\sigma$ takes, which means the obstacle encountered by Klein's algorithm is overcome.
%\begin{flushleft}
%\emph{with transition probability} $P(\mathbf{x}^l,\mathbf{x}^{l+1})=\frac{e^{-\frac{1}{2\sigma^2}\parallel \mathbf{Bx}^{l+1}-\mathbf{c} \parallel^2}}{\sum_{x_i^{l+1} \in \mathbb{Z}}e^{-\frac{1}{2\sigma^2}\parallel \mathbf{Bx}^{l+1}-\mathbf{c} \parallel^2}}$.
%\end{flushleft}
%\begin{flushleft}
%\emph{where the state space} $\Omega\in \mathbb{Z}^n$ \emph{is countable infinite}.
%\end{flushleft}
\newtheorem{my}{Lemma}

\subsection{Gibbs-Klein Sampling for Lattice Gaussian}

Although the afore-mentioned Gibbs sampler will converge to the stationary distribution eventually, the way it functions by individually sampling only one component each time leads to slow convergence. Especially, for lattice bases whose components are highly correlated with each other, the Markov chain induced by the standard Gibbs sampling can be trapped for a long time. To hasten convergence of the Markov chain, a new sampling algorithm combining Gibbs and Klein algorithms is proposed in the sequel.

%For a better illustration of the proposed sampling, here we establish another new scheme but equivalent to the foregoing one shown in Algorithm 2, which resorts to the help of permutation matrices.

The idea of blocked sampling is to sample a block of components of $\mathbf{x}$ at each step \cite{RobertsAndSahu}. Intuitively, this will lead to a faster convergence rate, which is already shown in \cite{LiuBook}. However, sampling a block is generally more costly than componentwise sampling. We propose to use Klein's algorithm for block sampling; this leads to the Gibbs-Klein.

At each step of the Markov chain, the proposed Gibbs-Klein sampling randomly picks up a block of $m$ components of $\mathbf{x}$ to update. For convenience, an $n \times n$ permutation matrix $\mathbf{E}$ is applied before blocking so that the blocks are updated in a fixed order.
%\begin{figure}[t]
%\begin{center}
%\begin{tikzpicture}
%\node  at (0,4) {\ \ \ $x^{t+1}_1$};
%\node  at (0,3) {$x^t_2$};
%\node  at (0,2) {$x^t_3$};
%\node  at (0,1) {$x^t_4$};
%%\node  at (0,0) {Standard};
%%\node  at (5,0) {Blocked};
%%\draw [thick](-1,0.5) --(1,0.5);
%%\draw [thick](4,0.5) --(6,0.5);
%\node  at (0,5) {Standard};
%\node  at (5,5) {Gibbs-Klein};
%\draw [->] (0.2,2) to [out=30,in=330] (0.2,4);
%\draw [->] (-0.2,1) to [out=130,in=220] (-0.2,4);
%\draw [->](0,3.2) --(0,3.8);
%\draw (1,2.55) --(2,2.55);
%\draw (1,2.45) --(2,2.45);
%\node  at (2,2.5) {$>$};
%\node  at (2.5,2.55) {\ $\mathbf{x}^{t+1}$};
%\draw (3,2.55) --(4,2.55);
%\draw (3,2.45) --(4,2.45);
%\node  at (3,2.5) {$<$};
%\node  at (5,4) {\ \ \ $x^{t+1}_1$};
%\node  at (5,3) {\ \ \ $x^{t+1}_2$};
%\node  at (5,2) {$x^t_3$};
%\node  at (5,1) {$x^t_4$};
%\draw [dashed] (4.5, 2.6) -- (5.6, 2.6);
%\draw [dashed] (4.5, 4.4) -- (5.6, 4.4);
%\draw [dashed] (4.5, 2.6) -- (4.5, 4.4);
%\draw [dashed] (5.6, 2.6) -- (5.6, 4.4);
%%\draw [->](5,2.2) --(5,2.8);
%\draw [->] (5.2,1) to [out=30,in=330] (5.72,3.5);
%\draw [->] (4.8,2) to [out=160,in=220] (4.35,3.5);
%\end{tikzpicture}
%\end{center}
%  \caption{Illustration of standard Gibbs sampling and Gibbs-Klein sampling. Components within the dashed block are sampled as a whole by Gibbs-Klein sampling.}
%  \label{Graphical representations}
%\end{figure}

Specifically, if $\mathbf{E}$ is random, then Gibbs-Klein sampling on $m$ randomly chosen components will be equivalent to sample $m$ consecutive components of $\mathbf{z}$ in a fixed order, where $\mathbf{z}=\mathbf{E^{-1}x}$ and $\mathbf{\widetilde{B}}=\mathbf{BE}$.
For simplicity, we always consider the block formed by the first $m$ components of $\mathbf{z}$, namely $\mathbf{z}_{\text{block}}=[z_1,\ldots,z_m]^T$. After QR-decomposition
$\mathbf{\widetilde{B}}=\mathbf{QR}$ and calculating $\mathbf{c}'=\mathbf{Q}^{T}\mathbf{c}$, $z_i$ in the block is sampled from the following 1-dimensional distribution with the backward order from $z_m$ to $z_1$:
%conditioned on other $n-m$ components $\mathbf{x}^{'}_{[-\text{block}]}=[x^{'}_{m+1},\ldots,x^{'}_n]^T$
\begin{eqnarray}
P(z_i^{t+1}|\mathbf{\overline{z}}_{[-i]}^t)&=&D_{\mathbb{Z},\alpha_i,\widetilde{z}_i^t},
\label{1 conditional distribution zzz}
\end{eqnarray}
where $\alpha_i=\frac{\sigma}{| r_{i,i}|}$, $ \mathbf{\overline{z}}_{[-i]}^t =[z_{i+1}^{t+1},\ldots,z_m^{t+1},z_{m+1}^t,\ldots,z_{n}^t]^T$ and $\widetilde{z}_i^t=\frac{c'_i-\sum^m_{j=i+1}r_{i,j}z_j^{t+1}-\sum^n_{j{'}=m+1}r_{i,j{'}}z_{j{'}}^t}{r_{i,i}}$. Algorithm 3 gives the proposed Gibbs-Klein sampling, where $\mathbf{z}^{t+1}=[\mathbf{z}^{t+1}_{\text{block}};\mathbf{z}^t_{[-\text{block}]}]$ is obtained after each step, and $\mathbf{z}^t_{[-\text{block}]}=[z_{m+1}^{t},\ldots,z_{n}^{t}]^T$. The implementation given in Algorithm 3 is not so efficient due to repeated QR decompositions; Optimizing for better efficiency will be pursued in the future. Note that the extension to other scan strategies is also possible.

\begin{algorithm}[t]
\caption{Gibbs-Klein sampling for lattice Gaussian}
\begin{algorithmic}[1]
\Require
$\mathbf{B}, \sigma, \mathbf{c}, m, \mathbf{x}^0$;
\Ensure
$\mathbf{x}$ from a distribution close to $D_{\Lambda,\sigma,\mathbf{c}}$ as $T \to \infty$
\For {$t=$1,\ \ldots,\ $T$}
\State randomly generate a permutation matrix $\mathbf{E}$
\State Let $\mathbf{\widetilde{B}}=\mathbf{BE}$ and $\mathbf{z}=\mathbf{E}^{-1}\mathbf{x}$
\State Let $\mathbf{\widetilde{B}}=\mathbf{QR}$ and $\mathbf{c}'=\mathbf{Q}^{T}\mathbf{c}$
\For {$i=m$,\ \ldots,\ 1}
\State let $\alpha_i=\frac{\sigma}{|r_{i,i}|}$
\State let $\widetilde{z}_i^{t-1}\hspace{-0.3em}=\hspace{-0.3em}\frac{c'_i-\sum^m_{j=i+1}r_{i,j}z_j^{t}-\sum^n_{j^{'}=m+1}r_{i,j^{'}}z_{j^{'}}^{t-1}}{r_{i,i}}$
\State sample $z_i^t$ from $D_{\mathbb{Z},\alpha_i,\widetilde{z}_i^{t-1}}$
%\State update $\mathbf{\overline{z}}_{[-(i-1)]}^{\ t-1}=[z_i; \mathbf{\overline{z}}_{[-i]}^{\ t-1}]$
\EndFor
\State update $\mathbf{z}^{t}=[\mathbf{z}^{t}_{\text{block}};\mathbf{z}^{t-1}_{[-\text{block}]}]^T$
%z_1^t,\ldots, z_m^t,z_{m+1}^{t-1},\ldots,z_{n}^{t-1}
\State return $\mathbf{x}^t=\mathbf{Ez}^t$
\If {Markov chain has reached stationarity}
\State output $\mathbf{x}^{t}$
\EndIf
\EndFor
\end{algorithmic}
\end{algorithm}

\section{Analysis of Gibbs-Klein sampling}
In this section, we show that the proposed Gibbs-Klein sampling algorithm can induce a reversible Markov chain within a negligible error. From (\ref{1 conditional distribution zzz}) and by induction, the sampling probability of $\mathbf{z}_{\text{block}}^{t+1}$ conditioned on $\mathbf{z}_{[-\text{block}]}^t$ is given by
\begin{equation}
P(\mathbf{z}^{t+1}_{\text{block}}\mid\mathbf{z}^t_{[-\text{block}]})=\prod^m_{i=1}P(z^{t+1}_{m+1-i}|\mathbf{\overline{z}}^t_{[-(m+1-i)]}).
\label{1 conditional distribution zzzz}
\end{equation}
%which considers multiple variate as a whole, which achieves a better convergence rate of Markov chain. Despite the depressed fact about the convergence rate or mixing time, we also give an upper bound about the mixing time of the Markov chain operated by the proposed Gibbs-sampling algorithm. For analytical simplicity, the theoretical analysis is performed on $\mathbf{z}$ rather than $\mathbf{x}$ while they are actually equivalent according to the application of $\mathbf{E}$.

%Let $\mathbf{z}_{\text{block}}=[z_1,\ldots,z_m]^T$ represent the vector contained in a block, then by induction from $z_m$ to $z_1$, the sampling probability over $\mathbf{z}_{\text{block}}$ can be derived as
%\begin{equation}
%P(\mathbf{z}_{\text{block}})=\prod^s_{i=1}P(z_{m+1-i}|\mathbf{\overline{z}}_{[-m-1+i]}).
%\label{1 conditional distribution zzzz}
%\end{equation}
%\begin{equation}
%P(\mathbf{z}_{\text{block}}^{l+1}|\mathbf{z}_{[-\text{block}]}^l)=\prod^j_{i=1}P(z_{s+1-i}^{l+1}|\mathbf{\overline{z}}_{[-s-1+i]}^l).
%\label{1 conditional distribution zzzz}
%\end{equation}

The following lemma gives a closed-form expression of this conditional probability within a negligible error and the proof follows \cite{Trapdoor}.
\begin{my}
For a given invariant distribution $D_{\Lambda,\sigma,\mathbf{c}}$, the transition probability $P(\mathbf{z}^{t+1}_{\text{block}}\mid\mathbf{z}^t_{[-\text{block}]})$ of Gibbs-Klein algorithm is within negligible statistical distance of the following distribution
\end{my}
\vspace{-1em}
\begin{equation}
D{'}=\frac{e^{-\frac{1}{2\sigma^2}\parallel \mathbf{\widetilde{B}z}^{t+1}-\mathbf{c} \parallel^2}}{\sum_{\mathbf{z}^{t+1}_{\text{block}} \in \mathbb{Z}^m}e^{-\frac{1}{2\sigma^2}\parallel \mathbf{\widetilde{B}z}^{t+1}-\mathbf{c} \parallel^2}}
\label{sampling radius bound}
\end{equation}
\begin{flushleft}
\emph{if} $\sigma \hspace{-0.3em}\geq\hspace{-0.3em} \omega(\hspace{-0.2em}\sqrt{\text{log}\hspace{-0.2em}\ m})\hspace{-0.2em}\cdot\hspace{-0.2em}\text{max}_{1\leq i\leq m}\hspace{-0.1em}\|r_{i\hspace{-0.1em},i}\hspace{-0.1em}\|$, \emph{where} $\mathbf{z}^{t\hspace{-0.1em}+\hspace{-0.1em}1}\hspace{-0.4em}=\hspace{-0.3em}[\mathbf{z}^{t\hspace{-0.1em}+\hspace{-0.1em}1}_{\text{block}}\hspace{-0.1em};\hspace{-0.1em}\mathbf{z}^t_{[-\text{block}\hspace{-0.1em}]}]$.
%\begin{my}
%For a given invariant distribution $D_{\Lambda,\sigma,\mathbf{c}}$, the conditional probability of Gibbs-Klein sampling returning a
%\end{my}
%\vspace{-1.1em}
%\begin{flushleft}
%\emph{candidate of} $\mathbf{z}^{t+1}_{\text{block}}\hspace{-0.3em}=\hspace{-0.3em}[\hspace{-0.1em}z_1^{t+1}\hspace{-0.05em},\hspace{-0.05em}\ldots\hspace{-0.05em},\hspace{-0.05em}z_m^{t+1}\hspace{-0.1em}]^T$ \emph{conditioned on} $\mathbf{z}_{[-\text{block}]}^t$ \emph{satisfies}
%\end{flushleft}
%\vspace{-1em}
%\begin{equation}
%P(\mathbf{z}^{t+1}_{\text{block}}\mid\mathbf{z}^t_{[-\text{block}]})\simeq \frac{e^{-\frac{1}{2\sigma^2}\parallel \mathbf{\widetilde{B}z}^{t+1}-\mathbf{c} \parallel^2}}{\sum_{\mathbf{z}^{t+1}_{\text{block}} \in \mathbb{Z}^m}e^{-\frac{1}{2\sigma^2}\parallel \mathbf{\widetilde{B}z}^{t+1}-\mathbf{c} \parallel^2}}
%\label{sampling radius bound}
%\end{equation}
%%\begin{equation}
%%\pi(\mathbf{z}_{\text{block}}^{l+1}|\mathbf{z}_{[-\text{block}]}^l)= \frac{e^{-\frac{1}{\alpha^2}\|{\underline{\mathbf{y}}}-\mathbf{\underline{R}}\mathbf{z}^{l+1}_{\text{block}}\|^2}}{\sum_{\mathbf{z}^{l+1}_{\text{block}}\in\mathcal{X}^s}e^{-\frac{1}{\alpha^2}\|{\underline{\mathbf{y}}}-\mathbf{\underline{R}}\mathbf{z}^{l+1}_{\text{block}}\|^2}}
%%\label{sampling radius bound}
%%\end{equation}
%\begin{flushleft}
%\emph{if} $\sigma \hspace{-0.1em}\geq\hspace{-0.1em} \omega(\sqrt{\text{log}\ m})\hspace{-0.1em}\cdot\hspace{-0.1em}\text{max}_{1\leq i\leq m}$
%$\|\mathbf{\widehat{b}}_i\|$, where ``$\simeq$" represents equality up to a negligible error.
\end{flushleft}
%\vspace{-0.5em}

\begin{proof}
According to (\ref{1 conditional distribution zzz}) and (\ref{1 conditional distribution zzzz}), we have
\begin{equation}
\hspace{-1.5cm}P(\hspace{-0.1em}\mathbf{z}_{\text{block}}^{t+1}\hspace{-0.3em}\mid\hspace{-0.3em}\mathbf{z}_{[-\text{block}]}^{t}\hspace{-0.1em})=\prod^m_{i=1}D_{\mathbb{Z},\alpha_{m\hspace{-0.1em}+\hspace{-0.1em}1\hspace{-0.1em}-\hspace{-0.1em}i},\widetilde{z}^t_{m\hspace{-0.1em}+\hspace{-0.1em}1\hspace{-0.1em}-\hspace{-0.1em}i}}(z_{m\hspace{-0.1em}+\hspace{-0.1em}1\hspace{-0.1em}-\hspace{-0.1em}i}^{t+1})\notag
\end{equation}
\vspace{-1em}
\begin{eqnarray}
\hspace{-0.5cm}&=&\hspace{-1.1em}\frac{e^{-\frac{1}{2\sigma^2}\sum_{i=1}^m\left(\overline{c}_{m\hspace{-0.1em}+\hspace{-0.1em}1\hspace{-0.1em}-\hspace{-0.1em}i}-\sum_{j=m\hspace{-0.1em}+\hspace{-0.1em}1\hspace{-0.1em}-\hspace{-0.1em}i}^mr_{m\hspace{-0.1em}+\hspace{-0.1em}1\hspace{-0.1em}-\hspace{-0.1em}i,j}z_j^{t+1}\right)^2}}{\prod^m_{i=1}\sum_{z_{m\hspace{-0.1em}+\hspace{-0.1em}1\hspace{-0.1em}-\hspace{-0.1em}i}^{t+1}\in \mathbb{Z}}e^{-\frac{1}{2\sigma^2}\left(\overline{c}_{m\hspace{-0.1em}+\hspace{-0.1em}1\hspace{-0.1em}-\hspace{-0.1em}i}-\sum_{j=m\hspace{-0.1em}+\hspace{-0.1em}1\hspace{-0.1em}-\hspace{-0.1em}i}^mr_{m\hspace{-0.1em}+\hspace{-0.1em}1\hspace{-0.1em}-\hspace{-0.1em}i,j}z_j^{t+1}\right)^2}}\notag\\
\hspace{-2.5cm}&=&\hspace{-1.1em}\frac{e^{-\frac{1}{2\sigma^2}\parallel\overline{\mathbf{c}}-\overline{\mathbf{r}}\mathbf{z}_{\text{block}}^{t+1}\parallel^2}}{\prod^m_{i\hspace{-0.13em}=\hspace{-0.13em}1}\hspace{-0.31em}\sum_{z_{m\hspace{-0.13em}+\hspace{-0.13em}1\hspace{-0.13em}-\hspace{-0.13em}i}^{t\hspace{-0.13em}+\hspace{-0.13em}1}\hspace{-0.23em}\in \mathbb{Z}}\hspace{-0.23em}e\hspace{-0.23em}^{-\hspace{-0.13em}\frac{1}{2\sigma^2}\hspace{-0.13em}\left(\hspace{-0.13em}r_{m\hspace{-0.13em}+\hspace{-0.13em}1\hspace{-0.13em}-\hspace{-0.13em}i,\hspace{-0.13em}m\hspace{-0.13em}+\hspace{-0.13em}1\hspace{-0.13em}-\hspace{-0.13em}i}z_{m\hspace{-0.13em}+\hspace{-0.13em}1\hspace{-0.13em}-\hspace{-0.13em}i}^{t\hspace{-0.13em}+\hspace{-0.13em}1}\hspace{-0.23em}-\hspace{-0.06em}\overline{c}_{m\hspace{-0.13em}+\hspace{-0.13em}1\hspace{-0.13em}-\hspace{-0.13em}i}\hspace{-0.13em}+\hspace{-0.13em}\sum_{j=m\hspace{-0.13em}+\hspace{-0.13em}2\hspace{-0.13em}-\hspace{-0.13em}i}^m\hspace{-0.23em}r_{m\hspace{-0.13em}+\hspace{-0.13em}1\hspace{-0.13em}-\hspace{-0.13em}i\hspace{-0.13em},\hspace{-0.13em}j}z_j^{t\hspace{-0.13em}+\hspace{-0.13em}1}\hspace{-0.13em}\right)\hspace{-0.13em}^2}}\notag\\
\hspace{-2.5cm}&=&\hspace{-1.1em}\frac{\rho_{\mathcal{L}(\overline{\mathbf{r}}),\sigma,\overline{\mathbf{c}}}(\mathbf{z}_{\text{block}}^{t+1})}{\prod^m_{i=1}\rho_{\sigma}(r_{m\hspace{-0.1em}+\hspace{-0.1em}1\hspace{-0.1em}-\hspace{-0.1em}i,m\hspace{-0.1em}+\hspace{-0.1em}1\hspace{-0.1em}-\hspace{-0.1em}i}\mathbb{Z}+\xi)},
\label{1 conditional distribution a}
\end{eqnarray}
where $\overline{c}_i=c'_i-\sum_{j^{'}=m+1}^nr_{i,j^{'}}z_{j^{'}}^{t}$, $\overline{\mathbf{c}}=[\overline{c}_1,\ldots, \overline{c}_m]^T$, $\xi=\sum_{j=m+2-i}^mr_{m+1-i,j}z_j^{t+1}-\overline{c}_{m-i+i}$ and $\overline{\mathbf{r}}$ is the $m \times m$ segment of $\mathbf{R}$ with $r_{1,1}$ to $r_{m,m}$ in the diagonal. Clearly, the effect of the subvector $\mathbf{z}_{[-\text{block}]}^t$ is hidden in $\overline{c}_i$.
In \cite{Regevlearning}, it has been demonstrated  that if $\sigma>\eta_\varepsilon(\mathcal{L}(\overline{\mathbf{r}}))$, then
%\vspace{0.2em}
\begin{equation}
\frac{\prod^m_{i=1}\rho_{\sigma}(r_{i,i}\mathbb{Z}+\xi)}{\prod^m_{i=1}\rho_{\sigma}(r_{i,i}\mathbb{Z})}\in\left(\left(\frac{1-\varepsilon}{1+\varepsilon}\right)^m,1\right]
\label{lemma 1}
\end{equation}
which means $\prod^m_{i=1}\rho_{\sigma}(r_{i,i}\mathbb{Z}+\xi)$ can be substituted by $\prod^m_{i=1}\rho_{\sigma}(r_{i,i}\mathbb{Z})$ within negligible errors when $\varepsilon$ is sufficiently small.

As shown in \cite{Trapdoor}, $\eta_\varepsilon(\Lambda)$ with negligible $\varepsilon$ is upper bounded as $\eta_\varepsilon(\Lambda)\leq \omega(\sqrt{\text{log}\ n})\cdot\text{max}_{1\leq i\leq n}\|\mathbf{\widehat{b}}_i\|$. Therefore, if $\sigma\geq \omega(\sqrt{\text{log}\ m})\cdot\text{max}_{1\leq i\leq m}\|r_{i,i}\|$, $P(\mathbf{z}_{\text{block}}^{t+1}\hspace{-0.3em}\mid\hspace{-0.3em}\mathbf{z}_{[-\text{block}]}^t)$ shown in (\ref{1 conditional distribution a}) can be rewritten as
\begin{equation}
P(\mathbf{z}_{\text{block}}^{t+1}\hspace{-0.3em}\mid\hspace{-0.3em}\mathbf{z}_{[-\text{block}]}^t)\simeq\frac{\rho_{\mathcal{L}(\overline{\mathbf{r}}),\sigma,\overline{\mathbf{c}}}(\mathbf{z}_{\text{block}}^{t+1})}{\prod^m_{i=1}\rho_{\sigma}(r_{i,i}\mathbb{Z})},
\end{equation}
where ``$\simeq$" represents equality up to a negligible error. Because the denominator is independent of $\mathbf{z}_{\text{block}}^{t+1}$, $\mathbf{z}_{[-\text{block}]}^{t}$ and $\mathbf{c}$, it can be viewed as a constant and the output has a lattice Gaussian distribution $D_{\mathcal{L}(\overline{\mathbf{r}}),\sigma,\overline{\mathbf{c}}}(\mathbf{z}_{\text{block}}^{t+1})$.
%Moreover, if the subvector $\mathbf{z}_{[-\text{block}]}^{t}$ hidden in $\overline{\mathbf{c}}$ is relaxed, it is easily checked that
%\begin{equation}
%D_{\mathcal{L}(\overline{\mathbf{r}}),\sigma,\overline{\mathbf{c}}}(\mathbf{z}_{\text{block}}^{t+1})= \frac{e^{-\frac{1}{2\sigma^2}\parallel \mathbf{\widetilde{B}z}^{t+1}-\mathbf{c} \parallel^2}}{\sum_{\mathbf{z}_{\text{block}}^{t+1} \in \mathbb{Z}^m}e^{-\frac{1}{2\sigma^2}\parallel \mathbf{\widetilde{B}z}^{t+1}-\mathbf{c} \parallel^2}},
%\end{equation}
%completing the proof.
\end{proof}

Then we arrive at the following proposition.

\begin{my3}
Suppose $\sigma \geq \omega(\sqrt{\log  m})\cdot\max_{1\leq i\leq m}\|\mathbf{\widehat{\tilde{b}}}_i\|$ at each step so that the negligible statistical distance is absorbed by numerical errors. Then, within numerical errors, the Markov chain induced by the Gibbs-Klein sampler is irreducible, aperiodic and reversible (hence positive recurrent) and converges to the stationary distribution in the total variation distance as $t\to \infty$:
\begin{equation}
\underset{t\rightarrow\infty}{\lim}\|P^t(\mathbf{x}; \cdot)-D_{\Lambda,\sigma,\mathbf{c}}\|_{TV}=0
\end{equation}
for all states $\mathbf{x}\in\mathbb{Z}^n$.
\end{my3}

\begin{proof}
%Because of $\mathbf{z}=\mathbf{E^{-1}x}$ and $\mathbf{\widetilde{B}}=\mathbf{BE}$, according to Lemma 1, we can easily derive that
%\begin{equation}
%P(\mathbf{x}^{t+1}_{\text{block}}\mid\mathbf{x}^t_{[-\text{block}]})\simeq \frac{e^{-\frac{1}{2\sigma^2}\parallel \mathbf{Bx}^{t+1}-\mathbf{c} \parallel^2}}{\sum_{\mathbf{x}^{t+1}_{\text{block}} \in \mathbb{Z}^m}e^{-\frac{1}{2\sigma^2}\parallel \mathbf{Bx}^{t+1}-\mathbf{c} \parallel^2}}
%\label{sampling radius bound a}
%\end{equation}
%if $\sigma \geq \omega(\sqrt{\text{log}\ m})\cdot\text{max}_{1\leq i\leq m}\|\mathbf{\widehat{b}}_i\|$, where $\mathbf{x}_{\text{block}}$ denotes $m$ random variate in $\mathbf{x}$ with respect to the other $n-m$ fixed components of $\mathbf{x}$.
Let $s_i$ and $s_j$ be two adjacent states in Gibbs-Klein sampling. For block size $m$, every two adjacent states in Gibbs-Klein sampling differ from each other by at most $m$ components.
For convenience, we express them as
\begin{eqnarray}
s_i=[\mathbf{x}_{\text{block}(i)},\mathbf{x}_{[-\text{block}]}] \hspace{0.6em}\text{and}\hspace{0.6em} s_j=[\mathbf{x}_{\text{block}(j)},\mathbf{x}_{[-\text{block}]}],
\end{eqnarray}
where $\mathbf{x}_{\text{block}(i)}$ and $\mathbf{x}_{\text{block}(j)}$ denote the $m$ components belonging to $s_i$ and $s_j$, respectively. Then, the transition probability of Gibbs-Klein sampling is
\begin{eqnarray}
P(s_i; s_j)&=&P(\mathbf{x}^{t+1}=s_j|\mathbf{x}^t=s_i) \notag\\
&=&P(\mathbf{x}^t_{\text{block}(i)}\rightarrow\mathbf{x}^{t+1}_{\text{block}(j)}|\mathbf{x}^t_{[-\text{block}]}) \notag\\
&\overset{(a)}{=}&P(\mathbf{x}^{t+1}_{\text{block}(j)}|\mathbf{x}^t_{[-\text{block}]})\notag\\
&\simeq&\frac{e^{-\frac{1}{2\sigma^2}\parallel \mathbf{B}s_j-\mathbf{c} \parallel^2}}{\sum_{\mathbf{x}^{t+1}_{\text{block}} \in \mathbb{Z}^m}e^{-\frac{1}{2\sigma^2}\parallel \mathbf{Bx}^{t+1}-\mathbf{c} \parallel^2}},
\label{bbbb}
\end{eqnarray}
where $(a)$ is due to the fact that $\mathbf{x}_{\text{block}}^{t+1}$ is sampled only conditioned on $\mathbf{x}_{[-\text{block}]}^t$.

%Therefore, we have
%\begin{equation}
%P(\mathbf{x}^{t+1}|\mathbf{x}^t)=P(\mathbf{x}^{t+1}_{\text{block}}\mid\mathbf{x}^t_{[-\text{block}]}).
%\end{equation}

To show the Markov chain is irreducible, we note that given a state $s$ one can attain with positive probability in one step
any state $s'$ which shares $>= (n-m)$ components with $s$. Now, if $s$ and $s'$ have, say, $d < n-m$ components in common, there is
always a positive probability that after each step they get exactly one more
component in common. So we can go in $n-d$ steps from one to the
other. But as soon as $m >= 2$, we can assume that at the first step
we get two more components in common, and then one at each further
step, so we can go with positive probability in $n-d-1$ steps.

%Because all states communicate with each other through the consecutive block moves, one can always go from a state to any other states although it may take more than one step. Therefore, there exists a positive integer $k$ such that
%\begin{equation}
%P(\mathbf{x}^{t+k}=s_j|\mathbf{x}^t=s_i)>0
%\end{equation}
%for any two states $s_i,s_j$, which means the Markov chain is irreducible.

On the other hand, it is clear to see that the number of steps required to move between any two states (can be the same state) is arbitrary without any limitation to be a multiple of some integer. Put another way, the chain is not forced into some cycle with fixed period between certain states. Therefore, the Markov chain is aperiodic.

As for reversibility, it is no hard to check that the following relationship holds
\begin{equation}
D_{\Lambda,\sigma,\mathbf{c}}(s_i)P(s_i;s_j)\simeq D_{\Lambda,\sigma,\mathbf{c}}(s_j)P(s_j;s_i)
\end{equation}
with the same expression
\begin{equation}
\frac{e^{-\frac{1}{2\sigma^2}\parallel \mathbf{B}s_i-\mathbf{c} \parallel^2}}{\sum_{\mathbf{x} \in \mathbb{Z}^n}e^{-\frac{1}{2\sigma^2}\parallel \mathbf{Bx}-\mathbf{c} \parallel^2}}\cdot\frac{e^{-\frac{1}{2\sigma^2}\parallel \mathbf{B}s_j-\mathbf{c} \parallel^2}}{\sum_{\mathbf{x}^{t+1}_{\text{block}} \in \mathbb{Z}^m}e^{-\frac{1}{2\sigma^2}\parallel \mathbf{Bx}^{t+1}-\mathbf{c} \parallel^2}},
\end{equation}
within negligible errors. Thus, the conclusion follows, completing the proof.
\end{proof}

%\begin{my1}
%Given the invariant distribution $D_{\Lambda,\sigma,\mathbf{c}}$, the Markov chain based on the proposed Gibbs-Klein sampler is positive recurrent and converges to the stationary distribution for all state spaces $\mathbf{x}\in\mathbb{Z}^n$ as follows
%\end{my1}
%\vspace{-1em}
%\begin{equation}
%\underset{t\rightarrow\infty}{\text{lim}}\|P^t(\mathbf{x},\cdot)-D_{\Lambda,\sigma,\mathbf{c}}\|_{TV}=0
%\end{equation}
%\begin{flushleft}
%\emph{if} $\sigma \geq \omega(\sqrt{\text{log}\ m})\cdot\text{max}_{1\leq i\leq m}\|\mathbf{\widehat{b}}_i\|$.
%\end{flushleft}

The advantages of Gibbs-Klein sampling are two-fold: compared with the conventional Gibbs sampling which only processes a single variate each time, it is more efficient to sample multiple variates in a block, improving the convergence rate; on the other hand, it overcomes the limitation of Klein's sampling which requires large values of $\sigma$ and extends lattice Gaussian sampling to the more general case.

\section{Simulation Results}
In this section, the performances of various sampling schemes are exemplified in the context of MIMO decoding. Specifically, we examine the decoding error probabilities to assess the convergence rates. By sampling from $D_{\Lambda,\sigma,\mathbf{c}}$, the closest lattice point will be returned with the highest probability, which implies an effective approach to lattice decoding.

\begin{figure}[t]
\includegraphics[width=3.5in]{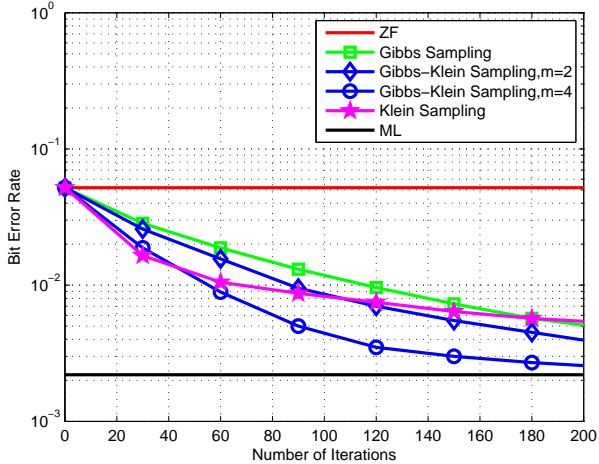}
\vspace{-1em}
  \caption{Bit error rate versus the number of iterations for the uncoded $4 \times 4$ MIMO system using 16-QAM.}
  \label{simulation 1}
\end{figure}

Fig. \ref{simulation 1} depicts the bit error rates (BER) of different Gibbs samplers in a $4\times4$ uncoded MIMO system with 16-QAM. This corresponds to lattice dimension $n=8$. The performances of zero-forcing (ZF) and maximum-likelihood (ML) decoding are also shown as benchmarks. We assume a flat fading environment with fixed SNR ($E_b/N_0\hspace{-0.3em}=\hspace{-0.3em}15$ dB). The channel matrix $\mathbf{H}$ consists of uncorrelated complex Gaussian fading gains with unit variance. $\mathbf{Hx}$ can be viewed as a lattice point in lattice $\Lambda\hspace{-0.2em}=\hspace{-0.2em}\mathcal{L}(\mathbf{H})$ and detecting the transmitted signal $\mathbf{x}$ corresponds to solving the CVP. Due to the finite constellation size, the implementation for discrete Gaussian sampling given in \cite{CongRandom} is followed.

Klein chose $\sigma=\text{min}_{1\leq i\leq n}\|\mathbf{\widehat{b}}_i\|/\sqrt{\text{log}\ n}$ and derived polynomial complexity $O(n^{\|\mathbf{Bx}-\mathbf{c}\|^2/\text{min}_i\|\mathbf{\widehat{b}}_i\|^2})$ for his algorithm to find the closest lattice point when it is not far from $\mathbf{c}$ \cite{Klein}. His derivation is essentially based on the assumption of a Gaussian distribution. However, we now know this choice of $\sigma$ does not satisfy the smoothing condition and thus his sampler does not really produce Gaussian samples \cite{Trapdoor}.

Here, we follow Klein's choice of $\sigma$ and apply the proposed Gibbs and Gibbs-Klein samplers to produce Gaussian samples from the lattice. For a fair comparison, when the block size is $m$, we run block sampling for $n/m$ times, and count this as a full iteration. This corresponds to one run of Klein's original algorithm which samples $n$ components.
As shown in Fig. \ref{simulation 1}, the decoding performance of all the sampling schemes improve with the number of iterations. With the same number of iterations (hence the same complexity), the decoding performance improves with the block size, which implies a faster convergence rate.

%\section{Conclusions}
%In this paper, we introduced MCMC methods into lattice Gaussian distributions to address the issue of small $\sigma$, where the useful way to sample from lattice Gaussian distributions when $\sigma < \omega(\sqrt{\text{log}\ n})\cdot\text{max}_{1\leq i\leq n}\|\mathbf{\widehat{b}}_i\|$ is lacking. To solve this problem, we firstly proposed a Gibbs sampling algorithm by building a reversible Markov chain. For the sake of faster convergence, we then proposed another sampling scheme known as Gibbs-Klein sampling to perform sampling. According to the analysis, a reversible Markov chain induced by Gibbs-Klein sampling can be established with a negligible error, which offers great flexibility on $\sigma$ by adjusting the block size.

\section*{Acknowledgment}
The authors would like to thank Damien Stehl\'{e} for helpful discussions.

\bibliographystyle{IEEEtran}
\bibliography{IEEEabrv,reference1}

\end{document}